\documentclass[authoryear,review]{elsarticle}
\makeatletter
\def\ps@pprintTitle{%
	\let\@oddhead\@empty
	\let\@evenhead\@empty
	\def\@oddfoot{}%
	\let\@evenfoot\@oddfoot}
\makeatother

\usepackage[utf8]{inputenc}
\usepackage{fontenc}[t1]

\usepackage{amsmath,amssymb,amsfonts}
\usepackage{mathtools}
\usepackage{fullpage}
\usepackage {cancel}

\usepackage{amsmath,amsfonts,amssymb}
\usepackage{geometry} 
\usepackage{graphicx}
\usepackage{epstopdf}
\usepackage{booktabs} 
\usepackage{array} 

\newcommand{\field}[1]{\mathbb{#1}}
\newcommand {\R}{\field{R} }

\newcommand{\G}{{\mathcal G}}
\newcommand{\A}{{\mathcal A}}

\newcommand {\ol}{\overline }
\newcommand {\e}{{\mathbf e} }
\newcommand {\hx} {{\hat x}}
\newcommand {\hy} {{\hat y}}
\newcommand {\tx} {{\tilde x}}
\newcommand {\ty} {{\tilde y}}

\newtheorem{prop}{Proposition}[section]
\newtheorem{ddef}[prop]{Definition}

\newtheorem{theorem}[prop]{Theorem}

\newtheorem{cor}[prop]{Corollary}

\usepackage{color}
\usepackage{colordvi}
\definecolor{magenta}{rgb}{.5,0,.5}
\definecolor{black}{rgb}{1.0,1.0,1.0}
\definecolor{magenta}{rgb}{.1,0,.3}
\definecolor{gruen}{rgb}{0.2,0.5,.5}
\definecolor{light}{rgb}{ 0.992, 0.961,  0.902}
\definecolor{Tan}{rgb}{ 0.992, 0.9,  0.902}

\newcommand{\komment}[1]{{}}

\usepackage{amsmath,amssymb}

\usepackage{changepage}

\usepackage{textcomp,marvosym}

\usepackage{amsmath}
\usepackage{amssymb} 

\usepackage{nameref,hyperref}

\usepackage{color}
\usepackage{colordvi}

\usepackage{microtype}
\DisableLigatures[f]{encoding = *, family = * }

\usepackage[table]{xcolor}

\usepackage{array}

\newcolumntype{+}{!{\vrule width 2pt}}

\newlength\savedwidth

\begin{document}

\begin{frontmatter}
	
	\title{Are the better cooperators dormant or quiescent?}
	
	\author[TUMWeihen]{Thibaut Sellinger} 
	\author[TUMGarching,ICB]{Johannes M\"uller\corref{mycorrespondingauthor}} 
	\author[TUMGarching]{Volker H\"osel} 
	\author[TUMWeihen]{Aur\'elien Tellier}

	\address[TUMGarching]{Center for Mathematics, Technische Universit\"at M\"unchen, 85748 Garching, Germany}
	\address[TUMWeihen]{Section of Population Genetics, Center of Life and Food Sciences Weihenstephan, Technische Universit\"at M\"unchen, 85354 Freising, Germany}
	\address[ICB]{Institute for Computational Biology, Helmholtz Center Munich, 85764 Neuherberg, Germany}

	\cortext[mycorrespondingauthor]{Corresponding author}


\begin{abstract}
Despite the wealth of empirical and theoretical studies, the origin and maintenance of cooperation is still an evolutionary riddle. In this context, ecological life-history traits which affect the efficiency of selection may play a role, though these are often ignored. We consider here species such as bacteria, fungi, invertebrates and plants which exhibit resting stages in the form of a quiescent state or a seedbank. When quiescent, individuals are inactive and reproduce upon activation, while under seed bank parents produce offspring remaining dormant for different amount of time. We assume weak frequency-dependent selection modeled using game-theory and the prisoners dilemma (cooperation/defect) as payoff matrix. The cooperators and defectors are allowed to evolve different quiescence or dormancy times. 
By means of singular perturbation theory we reduce the model to a one-dimensional equation resembling the well known replicator equation, where the gain functions are scaled with lumped parameters reflecting the time scale of the resting state of the cooperators and defectors. If both time scales are identical cooperation cannot persist in a homogeneous population. If, however, the time scale of the cooperator is distinctively different from that of the defector, cooperation may become a locally asymptotically stable strategy. Interestingly enough, in the seedbank case the cooperator needs to be faster than the defector, while in the quiescent case the cooperator has to be slower. We use adaptive dynamics to identify situations where cooperation may evolve and form a convergent stable ESS. We conclude by highlighting the relevance fo these results for many non-model species and the maintenance of cooperation in microbial, invertebrate or plant populations.
\end{abstract}
	
\end{frontmatter}


\section{Introduction}

After years of intensive research, the question how evolutionary forces generate and stabilize cooperation is still not fully understood. In a naive setting, cooperation is prone to exploitation of defecting individuals. The defectors have an advantage in e.g.\ feeding on a public  good produced by cooperators, which alone carry the costs for the public good production. Thus, the defectors are able to grow faster than the cooperators, and eventually out-compete them. This conflict between group- and personal benefit is often termed the tragedy of the commons~\citep{WestGriffin2006,AsfahlSchuster2016}. In a homogeneous population, cooperation will get lost~\citep{Hofbauer1998}. Cooperation among higher animals which have individuality can be based on direct and indirect reciprocal altruism~\citep{Trivers1971,schino2010,Nowak2006}. In other cases, as in bacteria or plants, most approaches that explain the persistence of cooperation and altruism are based on multilevel selection and assortmentn~\citep{price, Okasha2009}. 
In these cases, the population is  not homogeneous, but partitioned in locally interacting subunits. Assortment of cooperators counteract the tragedy of the commons. In homogeneous populations it is, however, difficult to find simple mechanisms that allow for cooperation to develop or to persist. Kin selection, the ability of cooperators  to recognize other cooperating individuals and to support only them, is one way out~\citep{West2007,West2007a,WestGardner2010}. Stable mechanisms for kin selection (that are not based on multilevel selection and assortment) are difficult. An approach is a ``greenbeard''-gene, a marker gene for cooperators~\citep{WestGardner2010}. Cooperators will be only responsive to greenbeards. This marker gene may be faked by ``falsebeards'' that carry some marker resembling the greebeard-marker and are therefore accepted by cooperators~\citep{WestGardner2010}. At least partially, the success of quorum sensing relies on kin selection~\citep{AsfahlSchuster2016}. Another idea is to punish non-cooperators. For example, the use of bactericides can be considered as such a strategy~\citep{WestGardner2010}. Smith introduced an elegant idea: plasmids could be carrier of cooperating genes, and defectors are infected with these plasmids. In this way, defectors become cooperators~\citep{Smith2001}. Nevertheless, there is also a  counter-argument: if plasmids can mutate and become defector-plasmids, this lead to destroy the cooperation~\citep{Ginty2010}. \par\medskip 

Using the same idea as the inhomogeneity effect on maintenance of cooperation, in a homogeneous population, we propose conceptually that cooperation is maintained if sufficient uncoupling occurs in time in the reproduction and competition between cooperators and defectors. The aim of the present paper is to explore the effect of such inhomogeneous time scales within a given population. Many species do exhibit for example resting stages as life-history traits. Many bacteria hibernate in a persister or sporulation state. In that state, the metabolism is low, such that the persister cells will not reproduce, but they are also less susceptible to antibiotics~\citep{Balaban2004}. Another form of hibernating states are seeds that rest in the soil~\citep{Honnay2008}. Both strategies can be considered as bet-hedging strategies that allow the population to survive an unpredictable and hostile time period~\citep{evans2005,Beaumont2009,lennon2011}. 
In this study, we investigate the interaction between cooperation and hibernation. We first develop an ecological model for the two situations, in the spirit of the replication equation~\citep{Hofbauer1998}. Two strategies are present in the population, and the two types (cooperators/defectors) are allowed to have a different resting stage strategy, \textit{i.e.} time scale. The model in the present case has not only two components (the active populations) but four (also the resting population components have to be taken into account).  The population growth, however, is assumed to be the same for both types. This assumption allows to reduce the system by singular perturbation theory to a one-dimensional equation. The resulting equation has a remarkable similarity to the replicator equation - only two new constants appear, that reflect the time scales of the two types. We discuss the circumstances that allow cooperation to persist in this system. \\
This first part focus on the ecological time scale. In the second part, we use adaptive dynamics in order to investigate if evolution promotes cooperation. Here we consider the evolutionary time scale, where rare mutations and selection lead to change the parameters of the system. We find that the strategy to defect is always an evolutionary stable state. This strategy, however, can be convergent unstable, such that a small perturbation is able to start an evolutionary dynamics that increases cooperation in the system. Thereto, the time scale of the resting state has to change with the degree of cooperation. Interestingly enough, seedbanks and quiescence require completely different dependencies: the cooperating, quiescent individuals have to stay longer in the resting state, while cooperator seeds need to germinate earlier than their defector counterparts.

\section{Model}

\begin{table}
\begin{center}
\begin{tabular}{l|l|l}
name & meaning (quiescence) & meaning (seedbank)\\
\hline
$x_i$ & active population &above ground population\\
$y_i$ & resting population & seed population\\
$N=x_1+x_2$ & total active pop. & total above ground pop.\\
$\beta_i$ & reproduction rate & seed production rate\\
$\alpha_i$ & become quiescent & -\\
$\mu_{x,i}$ & death (active indiv.) & death (above ground)\\
$\mu_{y,i}$ & death (resting indiv.) & death (seeds)\\
$\gamma_i$ & become active & germinate\\
$\omega $ & small parameter (weak selection) & small parameter (weak selection)\\
$g(.,.)$ & frequency dependence & frequency dependence\\
$\Pi$ & payoff matrix & payoff matrix\\
$c$ & costs for cooperation & costs for cooperation\\
$b$ & cooperation benefit& cooperation benefit
\end{tabular}
\end{center}
\caption{Meaning of variables and parameters in the quiescence and the seedbank model.}
\label{paraTab}
\end{table}

We discuss a model for quiescence and one for seedbanks in parallel, as these as well as the analysis closely resembles each other. Lengthy derivations are deferred to the appendix. In both models we describe the amount of individuals of two different types (or genotypes) interacting in the population. The types are found in an active subpopulation (metabolic active population for quiescence, and above-ground population for seedbanks). We denote this active subpopulation by $x_i$ ($i=1,2$). The quiescent population/seeds are denoted by $y_i$. We aim at a replicator equation~\citep{Hofbauer1998}. Therefore we choose our basic model to be linear (all individuals are independent, in particular, no competition). In reproduction, a weak interaction between the two types takes place. That is, the reproduction rate $\beta_i$ of type $i$ is modified by an additive term $\omega g_i(x_1,x_2)$ that expresses the effect of frequency-dependent selection on type~$i$. As usual for weak selection, $\omega$ is small. It is later possible to use $\omega$ as a small parameter in a singular perturbation approach and to reduce the dimension of the system.  \\
In the quiescence model, the active individuals in the population can reproduce, becomes resting (quiescent), or die. The quiescent individuals become active (wake up) at a certain rate while death also occurs (parameters are found in table~\ref{paraTab}). The model equations read
\begin{eqnarray}
\dot x_i &=& (\beta_i + \omega g_i(x_1,x_2))\, x_i - \alpha_i x_i-\mu_{x,i} x_i+\gamma_i y_i \label{modQx}\\
\dot y_i &=&  \alpha_i x_i-\gamma_i y_i -\mu_{y,i} y_i \label{modQy}
\end{eqnarray}
In the seedbank model, only the above-ground population reproduces. The offspring are seeds which can enter the seedbank and remain dormant for some amount of time depending the seed germination rate. The above-ground individuals and the seeds die at specific rates. We have
\begin{eqnarray}
\dot x_i &=&  \gamma_i y_i-\mu_{x,i} x_i \label{modSx}\\
\dot y_i &=&  (\beta_i + \omega g_i(x_1,x_2))\, x_i - \mu_{y,i} y_i -\gamma_i y_i\label{modSy}
\end{eqnarray}
We model the frequency-dependent interaction by a payoff-matrix $\Pi$ 
of a 2-player, 2 strategy game, as it is often done to address cooperation~\citep{Hofbauer1998}. We denote by $\e_i$ the $i$'th unit vector, and define the payoff of type $i$, 
$$ g_i(x_1,x_2) = \,\,\e_i^T \,\,\Pi\,\,
\left(\begin{array}{c}x_1\\x_2\end{array}\right)
\,\, \frac 1{x_1+x_2}$$
where this term is only defined if the total active population $N=x_1+x_2>0$. Later on we use the convention that $x_1$ denotes the population of cooperators, and $x_2$ that of defectors.

\subsection{Basic properties, projection, and neutral model}
We have $g_i(\zeta\,x_1, \,\zeta\,x_2)=g_i(x_1,x_2)$ if $\zeta>0$. That is, the model equations are homogeneous of degree one. Asymptotically, they tend to grow  exponentially~\citep{MuellerKuttler2015}. If $\omega=0$, the equations for the two types become uncoupled and the models become linear. Standard arguments show that the asymptotic exponentially growth rate $\lambda_i^{(q)}$ (for the quiescent model and genotype $i$) is given by the larger root of the characteristic equation
\begin{eqnarray}
p_q(\lambda^{(q)}) = (\lambda_i^{(q)})^2 - (\beta_i-\alpha_i-\mu_{x,i}-\gamma_i-\mu_{y,i}) \,\lambda_i^{(q)}\, - 
(\beta_i-\alpha_i-\mu_{x,i})(\gamma_i+\mu_{y,i})
-\alpha_i\gamma_i
=0,
\end{eqnarray}
while the exponent $\lambda_i^{(s)}$ for the seedbank model and genotype $i$ is  determined by the larger root of 
\begin{eqnarray}
p_s(\lambda^{(s)}) = (\lambda_i^{(s)})^2 + (\mu_{x,i}+\gamma_i+\mu_{y,i}  ) \,\lambda_i^{(s)}\, + 
\,\mu_{x,i}\,(\gamma_i+\mu_{y,i})
-\beta_i\gamma_i
=0
\end{eqnarray}
The exponents are always real as the matrices of the linear models are M-matrices~\citep{Berman2014}. 
\par\medskip

In order to further investigate the long term behavior, in particular for $\omega>0$,  we project the equations in defining new coordinates (recall that $N=x_1+x_2$) 
$$ \hx_i = x_i/N, \quad \hy_i = y_i/N.$$
Note that $\hx_1+\hx_2=1$, such that we only need an equation for $\hx_1$; if $\hx_2$ appears below, it always refers to $\hx_2=1-\hx_1$.\par\medskip 
We obtain for the quiescence model
\begin{eqnarray}
\frac d {dt} \hx_1  &=& 
(\beta_1 + \omega g_1(\hat x_1,\hat x_2))\, \hx_1 - \alpha_1 \hx_1
-\mu_{x,1} \hx_1+\gamma_1 \hy_1
-\hx_1\,\frac d {dt}\ln(N)
\label{qProjx}
\\
\frac d {dt} \hy_i  &=&  \alpha_i \hx_i-\gamma_i \hy_i -\mu_{y,i} \hy_i 
-\hy_i\,\frac d {dt}\ln(N)
\label{qProjy}\\
\frac d {dt}\ln(N) & = &
\sum_{i=1}^2\bigg( (\beta_i + \omega \, g_i(\hx_1,\hx_2)- \alpha_i -\mu_{x,i} )\hx_i+\gamma_i \hy_i\bigg)
\label{qProjN}
\end{eqnarray}
and for the seedbank model
\begin{eqnarray}
\frac d {dt} \hx_1  &=& - \mu_{x,1} \hx_1+\gamma_i \hy_1
-\hx_1\,\frac d {dt}\ln(N)
\label{sProjx}\\
\frac d {dt} \hy_i  &=& (\beta_i + \omega g_i(\hx_1,\hx_2))\, \hx_i - \mu_{y,i} \hy_i -\gamma_i \hy_i
\label{sProjy}
-\hy_i\,\frac d {dt}\ln(N)\\
\frac d {dt} \ln(N) &=& \sum_{i=1}^2\bigg(- \mu_{x,i}  \hx_i+\gamma_i \hy_i\bigg)
\label{sProjN}
\end{eqnarray}

In the following, let always $\ast\in\{q,s\}$, depending on the model considered 
(quiescence/seedbank). 
As we know that for $\omega=0$ the subpopulations for types~1 and~2 do not 
interact and grow exponentially, we find:
\begin{cor}
For $\omega>0$, $\omega$ small enough, $\hx_1$, $\hy_1$ tend to zero and $\hx_2$ 
tends to $1$ if $\lambda_2^{\ast}>\lambda_1^{\ast}$. If, in turn,  $\lambda_2^{\ast}<\lambda_1^{\ast}$, then $\hx_1$ tends to one while $\hx_2$, $\hy_2$ tend to zero.
\end{cor}
That is, the (weak) frequency dependent selection only has an influence on the dynamics if $\lambda_1^{(\ast)}=\lambda_2^{(\ast)}$. In this case, both types grow (for $\omega=0$) at the same rate. We are inclined to talk about a neutral model in this case. 

\begin{ddef}
We call the model neutral if $\lambda_1^{(\ast)}=\lambda_2^{(\ast)}$.
\end{ddef}

For the neutral model, we drop the index (as $\lambda_1^{(\ast)}=\lambda_2^{(\ast)}$) and just write $\lambda^{(\ast)}$. Also the next proposition supports our notion of neutrality: A line of stationary points appears. The fitness of the two types are completely equivalent, such that they persist in arbitrary relative frequencies. 
This line is transversally stable, that is, any trajectory with positive initial conditions eventually tends to a point on this line.

\begin{prop}\label{stableLineOfEquilibria}
If $\omega=0$ and the models are neutral, there is a line of stationary points in the projected models that is transversally stable. This line is given by $\hx_1\in[0,1]$, $\hx_2=1-\hx_1$, and $\hat y_i=\hat y_i^{equi}(\hx_1)$ with \\
Quiescence model:
\begin{eqnarray}
\hy_i^{equi}(\hx_1) :=  \frac{-\beta_i  + \alpha_i 
	+\mu_{x,i} +\lambda^{(q)} }{\gamma_i}\,\,\hat x_i
\end{eqnarray}
Seedbank model:
\begin{eqnarray}
\hy_i^{equi}(\hx_1) := \frac{\mu_{x,i}+\lambda^{(s)} }{\gamma_i  }\,\, \hx_i
\end{eqnarray}
\end{prop}
{\bf Proof:} For $\omega=0$, both subpopulations $(x_i, y_i)$ 
evolve independently, and tend to an exponentially growing 
population. We consider eqn.~(\ref{qProjx}) resp.~(\ref{sProjx}) to infer $\hy_i$ in terms of $\hx_i$ (where we take into account that $\frac d {dt}\ln(N)$ is -- in the exponentially growing solution -- just the corresponding exponent). Let us focus on the first subpopulation. If the initial value is larger zero, we find its limiting 
behavior characterized by the fraction of active and resting population, given by some variables $\check x_1$, $\check y_1\geq 0$, $\check x_1+\check y_1=1$ and $c_1>0$ such that 
$$ e^{-\lambda^{(\ast)}\, t}(x_1(t),y_1(t)) \rightarrow 
c_1\, (\check x_1,\check y_1).$$
The same for population~2.  
Hence, for $t\rightarrow\infty$, 
$$ \tilde x_i(t)=\frac{x_i(t)}{x_1(t)+x_2(t)}\rightarrow \frac {c_i}{c_1+c_2}\,\,\check x_i,
\quad 
 \tilde y_i(t)\rightarrow \frac {c_i}{c_1+c_2}\,\,\check y_i.$$
 The solution tends in the long run to the line of stationary points.\qed\par\medskip

\section{Weak selection}

If $\omega>0$ but small, the singular perturbation theory~\citep{OMalley2012} provides the appropriate framework to discuss the dynamics. We expect a solution to converge ${\cal O}(\omega)$-close to the line of stationary points on a fast time scale, and consequently to move slowly on an invariant manifold, again ${\cal O}(\omega)$-close to the line of stationary points. We are in particular after this slow long term behavior. As the long-term dynamics is expected to stay $\omega$-close to the line of invariant points, we introduce new coordinates $(\tx_1,\ty_1,\ty_2)$,
\begin{eqnarray}
\tx_1(t) = \hx_1(t),\qquad \hy_i(t) = \hy_i^{equi}(\hx_1(t))+\omega \ty_i(t).
\end{eqnarray}
and the abbreviation $\tx_2=1-\tx_1$. 
\begin{prop} 
The neutral quiescence model for $\omega>0$ reads (defining $\ol i=2$ if $i=1$ and {\it vice versa})
\begin{eqnarray}
\frac d{dt}\tx_i &=&  \omega \,\left[ \tx_i\,(1-\tx_i)\, 
\bigg\{ 
g_i(\tx_1,\tx_2)- g_{\ol i}(\tx_1,\tx_2)\bigg\}
+ \, \bigg\{
\gamma_i\,(1-\tx_i)\ty_i-\gamma_{\ol i}\,\tx_i\,\ty_{\ol i}\bigg\}\right]\\
	 \frac d {dt}\ty_i 
	&=& 
\frac{\beta_i  - \alpha_i 
	-\mu_{x,i} -\lambda^{(q)} }{\gamma_i}\,\,
\,\,\, \tx_i\,g_i(\tx_1,\tx_2)\\
&&\qquad+  
\bigg(\beta_i  - \alpha_i 
-\mu_{x,i} -\lambda^{(q)}-\gamma_i-\mu_{y,i}-\frac d {dt}\ln(N)\bigg) 
\,\,\ty_i \nonumber\\
	\frac d {dt}\ln(N) 
& = &  \lambda^{(q)} + \omega \sum_{j=1}^2\bigg(   g_j(\hx_1,\hx_2)\tx_j +\gamma_j\ty_j   \bigg)
\end{eqnarray}
\end{prop}
Proof:~\ref{QtransForm}.

\begin{prop} 
	The neutral seed bank model for $\omega>0$ reads 
	(defining $\ol i=2$ if $i=1$ and {\it vice versa})
	\begin{eqnarray}
\frac{d}{dt}\tx_{i}
&=&\omega\bigg(\gamma_{i}\,(1-\tx_i)\,\ty_{i}-\gamma_{\ol i}\,\tx_{i}\ty_{\ol i}\bigg)\\
\frac{d}{dt}\ty_{i}
&=&
g_i(\tx_1,\tx_2)\, \tx_i
- \left(\mu_{y,i} +\gamma_i +\mu_{x,i}+\lambda^{(s)} + \frac d {dt}\ln(N)   \right)\,\ty_i\\
\frac{d}{dt}ln(N)
&=&
\lambda^{(s)}+ \omega\,\sum_{i=1}^{2}\gamma_{i}\,\ty_{i}
	\end{eqnarray}
\end{prop}
Proof:~\ref{StransForm}.\par\medskip 
Note that the models are now in the standard form for singular perturbation
theory. We first consider the fast time scale, taking $\omega\rightarrow 0$. Then, $\dot\tx_i=0$, 
$\frac d {dt}ln(N)=\lambda^{(\ast)}$, and the equation for 
$\ty_i$ becomes linear. As the constants of the linear equation 
are negative (\ref{qInequlaities} for the quiescence model,~\ref{qInequlaitiesB} for the seedbank model), 
$\ty_i$ saddle fast on their stationary point, 
$$ \ty_i^{(\ast)} = A_i^{(\ast)} g_i(\tx_1,\tx_2)\,\tx_i/\gamma_i$$
where the lumped parameters $A_i^{(\ast)}$ read 
for the quiescence model
\begin{eqnarray}\label{Aseedbank}
A_i^{(q)} = -\,\, \frac{\beta_i-\alpha_i-\mu_{x,i}-\lambda^{(q)}}
	               {\beta_i-\alpha_i-\mu_{x,i}-\gamma_i-\mu_{y,i}-2\lambda^{(q)}}
\end{eqnarray}
while that for the seedbank model are given by 
\begin{equation}
A_i^{(s)} =\frac{\gamma_{i}}{\mu_{y,i} +\gamma_i +\mu_{x,i}+2\lambda^{(s)} }=1+ A_i^{(s')}, \qquad 
 A_i^{(s')}=-\frac{\mu_{x,i}+\mu_{y,i}+2\lambda^{s}}{\mu_{x,i}+\gamma_{i}+\mu_{y,i}+2\lambda^{s}}. 
\end{equation}
The numerator as well as denominator of (\ref{Aseedbank}) are negative (\ref{qInequlaities}) such that $A_i^{(q)}$ are negative. However, as the variables $\tilde y_i$ only indicate the first order deviation from the line of stationary points, they are allowed to be negative -- $\hat y_i$ are still positive, and the solution is biologically meaningful. In any case, $1+A_i^{(s')}$ and  $1+A_i^{(q)}$ are non-negative for $\lambda^{(\ast)}>0$. \par\medskip 

To obtain the dynamics on the slow manifold, we plug 
this quasi-stationary state into the equation of $\tx_1$, and 
find the following corollary.

\begin{cor} The dynamics of the reduced system on the  
	slow manifold in evolutionary time scale $T=\omega\, t$ 
	is given by 
\begin{eqnarray}
\frac d{dT} \tx_1 &=& 
\tx_1\, (1-\tx_1)\, 
\bigg\{
g_1(\tx_1,1-\tx_1)  \left(1+ A_1^{(\ast)}\,\right)
-
g_2(\tx_1,1-\tx_1) \left(1+ A_2^{(\ast)}\,\right)
\bigg\}.
\end{eqnarray}
\end{cor}

Note that we simply obtain a generalized replicator equation, where in contrast to the system without inner time scale, the gain functions $g_i(\tx_1,\tx_2)$ are modified by constants $A_i^{(\ast)}$ that are caused by the quiescent state respectively the seedbank. If $A_i^{(\ast)}=0$ we are back in the standard case~\citep{Hofbauer1998}. This observation is in line with previous findings indicating that models with seedbanks can be well approximated by models without seedbanking, but with rescaled parameters~\citep{Blath2013,Blath2015a,Koopmann2017,Heinrich2017}.

\subsection{ Prisoners dilemma and time scales}

We first note that the functions $g_i(\tx_1,\tx_2)$ are linear. Apart of the stationary states $\tx_1=0$ and $\tx_1=1$, there is at most one more interior  stationary state. In case that this interior stationary state does not exist, we find  directional selection, and any trajectory tends to a boundary equilibrium. If the interior stationary state is present, either it is unstable (disruptive selection, 
each boundary equilibrium is locally asymptotically stable), or locally stable (balancing selection, all trajectories in the interior tend to the interior stationary state).\par\medskip 

We investigate the prisoners dilemma~\citep{Hofbauer1998}, given by the payoff matrix 
$$ \mbox{Payoff-Matrix} = \left(\begin{array}{cc}
b-c & -c\\
b  & 0
\end{array}
\right)$$
where we assume for now that $b>c>0$. Hence, 
\begin{eqnarray}
g_1(\tx_1,\tx_2) 
= -c\,\tx_2\,+\,(b-c)\,\tx_1
=b\,\tx_1\,-\,c,\qquad 
g_2(\tx_1,\tx_2) 
= b\tx_1. 
\end{eqnarray}
The generalized replicator equation in that case reads 
$$ 
\frac d{dT} \tx_1  = 
\tx_1\, (1-\tx_1)\, 
\bigg\{
(b\,\tx_1-c) \left(1+ A_1^{(*)}\right)
-
b\tx_1 \left(1+ A_2^{(*)}\right)
\bigg\}.
$$

The sign of the curled bracket determines the stability of the equilibria. We obtain at once the following corollary. 

\begin{cor} 
In both model, the strategy to defect (stationary point $\tx_1=0$) is for all parameter choices locally asymptotically stable.  If 
$$ b \, \left(A_1^{(*)}-A_2^{(*)}\right) \leq c\, \left(1+A_1^{(*)}\right)$$
this strategy is even globally stable (directional selection). In the other case, 
\begin{eqnarray}
b \, \left(A_1^{(*)}-A_2^{(*)}\right) > c \,\left(1+A_1^{(*)}\right), 
\label{locStabCond}
\end{eqnarray}
the strategy to cooperate (stationary point $\tx_1=1$) is locally asymptotically stable. In this case, there is a unstable inner equilibrium (disruptive selection).\\ 
Balancing selection, \textit{i.e.} coexistence of cooperators and defectors in a locally asymptotically stable stationary point, cannot happen.
\end{cor}

Condition (\ref{locStabCond}) can never be satisfied for $A_i^{(\ast)}=0$. That is, in the classical replicator equation, cooperators cannot persist in the long run. 
Cooperation will also die out if the time scale of both types are equal ($A_1^{(\ast)}=A_2^{(\ast)}$). Only if quiescence resp.\ seedbanks are present in the system, and the time scale for the resting state is distinctively 
different for the two types such that the coefficient $A$ for the cooperator is larger than that for the defector, cooperation may become (locally) stable. 
\par\medskip

The basic assumption in the present theory is the neutrality of the two types. The time scale of the seedbank/quiescence state is in particular determined by $\gamma$. If we fix all parameters but $\alpha$ (quiescent model) resp.\ $\mu_x$ (seedbank model), the condition that $\lambda^{(\ast)}$ is fixed and constant superimposes a dependency of $\gamma$ and $\alpha$ (resp.\ $\mu_x$): If we vary $\gamma$, we are able to determine $\alpha=\alpha(\gamma)$ (resp.\ $\mu_x=\mu_x(\gamma)$), s.t.\ $\lambda^{(\ast)}$ stays constant. It is now possible to discuss the dependency of $A^{(\ast)}$ on the time scale of the seedbank/quiescent state, given by $\gamma$. 

\begin{prop} Let $\gamma_0>0$ and $\alpha_0>0$ ($\mu_{x,0}>0$) correspond to a growth rate $\lambda_0>0$. \\
Quiescence: $\alpha(\gamma)>0$ is well defined and non-negative for $\gamma>0$. $A^{(q)}=A^{(q)}(\gamma)$ is monotonousness decreasing in $\gamma$, and $A^{(q)}(\gamma)\in[0,-1]$.\\
Seedbank: There is a $\widehat\gamma_0>0$ such that $\mu_x(\gamma)$ is well defined and non-negative for $\gamma>\widehat \gamma_0$. $A^{(s')}=A^{(s')}(\gamma)$ is monotonousness increasing in $\gamma$, and $A^{(q)}(\gamma)\in[0,-1]$.\\
\end{prop}
The proof is given in appendix~\ref{quiescMonoRange} and~\ref{seedbankMonoRange}. 
All in all, in order to satisfy  $A_1^{(\ast)}-A_2^{(\ast)}>0$, we choose $A_i^{(\ast)}=A(\gamma_i)$, and $\gamma_1<\gamma_2$ for te quiescence model (resp. $\gamma_1>\gamma_2$ fr the seedbank model). The time scale of the defector ($\gamma_2$) needs to be faster (resp. slower) than that of the cooperator ($\gamma_1$).

The fitness of the population is only determined by the active population; in case of the 
quiescence model, we have 
$$\frac d {dt}\ln(x_i) = \beta_i+\omega g_i(x_1,x_2)-\alpha_i-\mu_{x,i}+\gamma_iy_i/x_i$$
while that for the seedbank model reads
$$\frac d {dt}\ln(x_i) = \gamma_iy_i/x_i-\mu_{x,i}.$$
In both cases, we find that the fitness is increased if he ratio $\gamma_i\,y_i/x_i$ (resting over active population, scaled by the rate to become active or to germinate, respectively) is increased. As $\gamma_i\,y_i/x_i=A_i^{(\ast)}\,g(\tilde x_1,\tilde x_2)$, this observation explains why large $A^{(\ast)}$ for the cooperators does favor their maintenance. The resting state acts an amplifier for (weak, frequency-dependent) selection effects.\par\medskip
In quiescence, reproduction goes to the active population. In order to increase the 
ratio of resting and active population (by the weak selection term) the resting population 
should become slower. In seedbanks, the reproduction directly supplies the below-ground 
population. In that case, seeds should germinate faster.

\section{Adaptive Dynamics}

Until now we mainly focused on the ecological time scale addressing the question of which type (cooperator and defector) out-competes the other in a population. We turn from ecology to evolution by assuming that the resting strategies of cooperators and defectors can now evolve over time. 
We assume that from time to time rare mutants appear in a population of 
residents (cooperators or defectors). If the mutants can spread and take over, evolution does a step: The parameters of the resident population changes. That is, while the dynamics of interactions within the population 
is interesting on the ecological time scale, the change of parameters (traits) themselves can be driven by evolutionary dynamics. The main tool to study such evolutionary change in our (deterministic) setting is Adaptive Dynamics~\citep{Geritz1996,DiekmannIntro}. 
As terms are sometimes used in slightly different ways, we recall some definitions. An unbeatable strategy (a strategy that cannot be invaded by other strategies, at least if they are close) is called an evolutionary steady state ESS. An ESS is called evolutionary convergent stable if Adaptive Dynamics predicts that strategies close-by are driven towards the ESS by evolutionary forces (here selection), and evolutionary unstable if evolutionary forces will drive strategies away.
\par\medskip 

In our setting, a trait is characterized by the time scale of the quiescent state/seedbank (represented by $A^{(\ast)}(\gamma)$), and the degree of cooperativeness, which is described by the generated benefit $b$. The production of 
benefits comes with costs for the producing individual, with larger benefits being more costly. The generic cost parameter is defined as $c=C(b)$ for some function $C$. The natural assumptions are $C(0)=0$ (no benefit, no cost), and $C'(b)\geq 0$ 
(costs are non-decreasing with the benefit value). Te trait of an individual is completely characterized by the tuple $(b, A^{(\ast)})$. 
\par\medskip

Under the assumption of trait independence, the evolution of resting state and cooperation would be uncoupled. However, genetic features such as genetic linkage of the genes underpinning the traits or genetic pleiotropy, or trade-off functions would link the evolution of both traits (resting stage and cooperation). Mutations at the underlying genes affect therefore $A^{(\ast)}$ and $b$. For simplicity, we assume here one single gene controlling cooperation and the time scale of the resting state (quiescence or dormancy), so that $A^{(\ast)}$ is a function of $b$ ($A^{(\ast)}=\A(b)$). It is thus sufficient to know $b$ to characterize the complete trait, $(b, \A(b))$. We want to compute the functions $\A(b)$ of quiescence/seedbank time scales and $C(b)$ of costs/benefits for which cooperation is favoured.\par\medskip

If the resident has trait $b_r$ and a time scale given by $A_r=\A(b_r)$, while that of the mutant has trait $b_m$ (time scale $A_m=\A(b_m)$), 
then the payoff matrix reads 
$$  \mbox{Payoff-Matrix} = \Pi = \left(\begin{array}{cc}
b_r-C(b_r) & b_m-C(b_r)\\
 b_r-C(b_m) & b_m-C(b_m)
 \end{array}\right).$$
If only the resident is present, then $x_1=1$. 
We aim to know two different properties of the mutant: 
(a) Can the mutant invade the resident? 
(b) Is the mutant able to outcompete the resident, and to take over the system?

\begin{prop}\label{invPropo} Define $A_r=\A(b_r)$, $A_m=\A(b_m)$, and 
$$\G(x_1; b_r, b_m) = 
  (1+A_r)\,  \bigg\{ C(b_m)-C(b_r)  \bigg\}
+ (A_r-A_m)\,\bigg\{  (b_r-b_m)\, x_1 + b_m-C(b_m) \bigg\}.
$$
Then, the generalized replicator equation reads 
$$ \dot x_1 = x_1 \, (1-x_1) \, \G(x_1; b_r, b_m).$$
The mutant can invade if $\G(1; b_r, b_m)<0$. The mutant can take over if $\G(1; b_r, b_m)<0$ and $\G(0; b_r, b_m)<0$.  
\end{prop}
{\bf Proof: }
We have 
\begin{eqnarray*}
g_1(x_1,x_2) &=& (b_r-C(b_r))\,x_1 + (b_m-C(b_r))\, x_2 
= (b_r-b_m)\,x_1  + b_m-C(b_r)\\
g_2(x_1,x_2) &=&   (b_r-C(b_m))\,x_1 + (b_m-C(b_m))\, x_2 
= (b_r-b_m)\,x_1+ b_m-C(b_m).
\end{eqnarray*}
The function $\G(x_1; b_r, b_m)$, as defined above, is nothing else 
but $\G(x_1; b_r, b_m)=g_1(x_1,x_2)(1+A_r)-g_2(x_1,x_2)(1+A_m)$, 
and hence the generalized replicator equation is given by 
$ \dot x_1 = x_1(1-x_1)\, \G(x_1; b_r, b_m)$. The remaining statements of this 
proposition are a direct consequence of the linearity of $\G$ in $x_1$.\qed\par\medskip 

Based on this proposition, we can (at least partially) answer our 
original question. If we assume that the resident has trait $b$, 
we find a criterion that mutant with (slightly/infinitesimal) 
larger $b$ are able to take over. 

\begin{prop} \label{infinitInvPropo}
If 
\begin{eqnarray}
 (b-C(b))\,\A'(b) \,>\, C'(b)\,  (1+\A(b)) \label{adapDynInequal}
 \end{eqnarray}
then Adaptive Dynamics predicts $b$ to increase under the pressure of evolutionary forces.
\end{prop}
{\bf Proof:} 
We show that the inequality given above implies $G(1;b,b+\varepsilon)<0$ and $G(0;b,b+\varepsilon)<0$ for $\varepsilon>0$, while for $\varepsilon<0$ the reversed inequalities hold true. Consider the Taylor expansion of $G(x_1;b,b+\varepsilon)$ w.r.t. $\varepsilon$. First of all, $\G(x_1; b,b)\equiv 0$. Next, the derivative of $\G(x_1;b,\ol b)$ w.r.t.\ $\ol b$ at $\ol b=b$ is given by 
\begin{eqnarray*}
\frac {\partial}{\partial \ol b}
\G(x_1;b,\ol b)\bigg|_{\ol b=b} 
&=& 
(1+\A(b))\,C'(b) \, - \, \A'(b)\,(b-C(b)) <0 
\end{eqnarray*}
where we used the assumption of the proposition in the last inequality. 
Note that the zero'th and the first order term are independent on $x_1$. 
As $G(x_1;b,b+\varepsilon) = \frac {\partial}{\partial \ol b}
\G(x_1;b,\ol b)\bigg|_{\ol b=b}\,\varepsilon + {\cal O}(\varepsilon^2)$, and the derivative is negative, the result follows. \qed\par\medskip

\begin{theorem}
The strategy $b=0$ cannot be invaded by any strategy $b>0$. 
If $C'(0)>0$, $b=0$ is evolutionary convergent stable. If $C'(0)=0$, $C''(0)$ sufficiently small,  
and $\A'(0)>0$, it is evolutionary unstable.
\end{theorem}
{\bf Proof: } 
First of all, 
$$  	\G(1;0,b_m) = (1+A_r)C(b_m)-(A_r-A_m)C(b_m) = (1+A_m)C(b_m)>0 \quad \mbox{ for } b_m>0$$
Proposition~\ref{invPropo} indicates that no mutant with $b_m>0$ can invade the strategy $b_r=0$. The strategy $b=0$ is an ESS. Furthermore, for $b=0$, we have $b-C(b)=0$, and hence 
\begin{eqnarray*}
	\frac {\partial}{\partial \ol b}
	\G(x_1;b,\ol b)\bigg|_{\ol b=b=0} 
	&=& 
	(1+\A(0))\,C'(0).
\end{eqnarray*}
If $C'(0)>0$, proposition~\ref{infinitInvPropo} shows that the ESS is evolutionary convergent stable. 
Let us assume that $C'(0)=0$ and ${\mathcal A}'(0)>0$. Then $C(b)=\frac 1 2 C''(0)b^2+{\cal O}(b^3)$, 
and  $C'(b)=C''(0)b+{\cal O}(b^2)$, s.t.\ 
$$ (b-C(b))\,\A'(b) \,-\, C'(b)\,  (1+\A(b))
= b(\A'(0)-C''(0)(1+\A(0)))+{\cal O}(b^2).$$ 
Hence for $b$ and $C''(0)$ sufficiently small but positive we have 
$(b-C(b))\,\A'(b) \,>\, C'(b)\,  (1+\A(b))$. Proposition~\ref{infinitInvPropo} 
proves that the ESS $b=0$ is evolutionary unstable.\qed\par\medskip

\begin{figure}
	\begin{center}
(a)	\includegraphics[width=6cm]{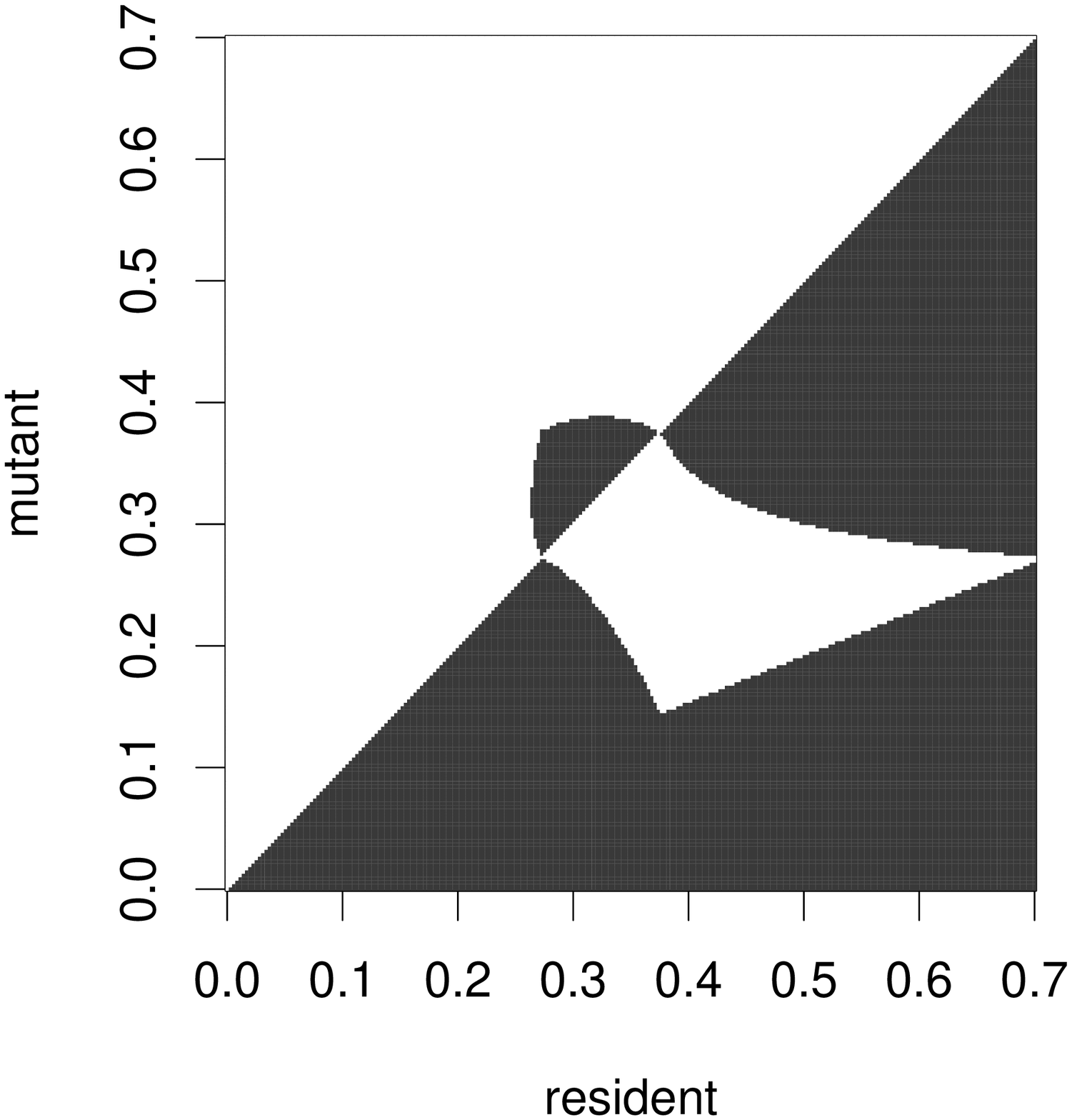}
(b)	\includegraphics[width=6cm]{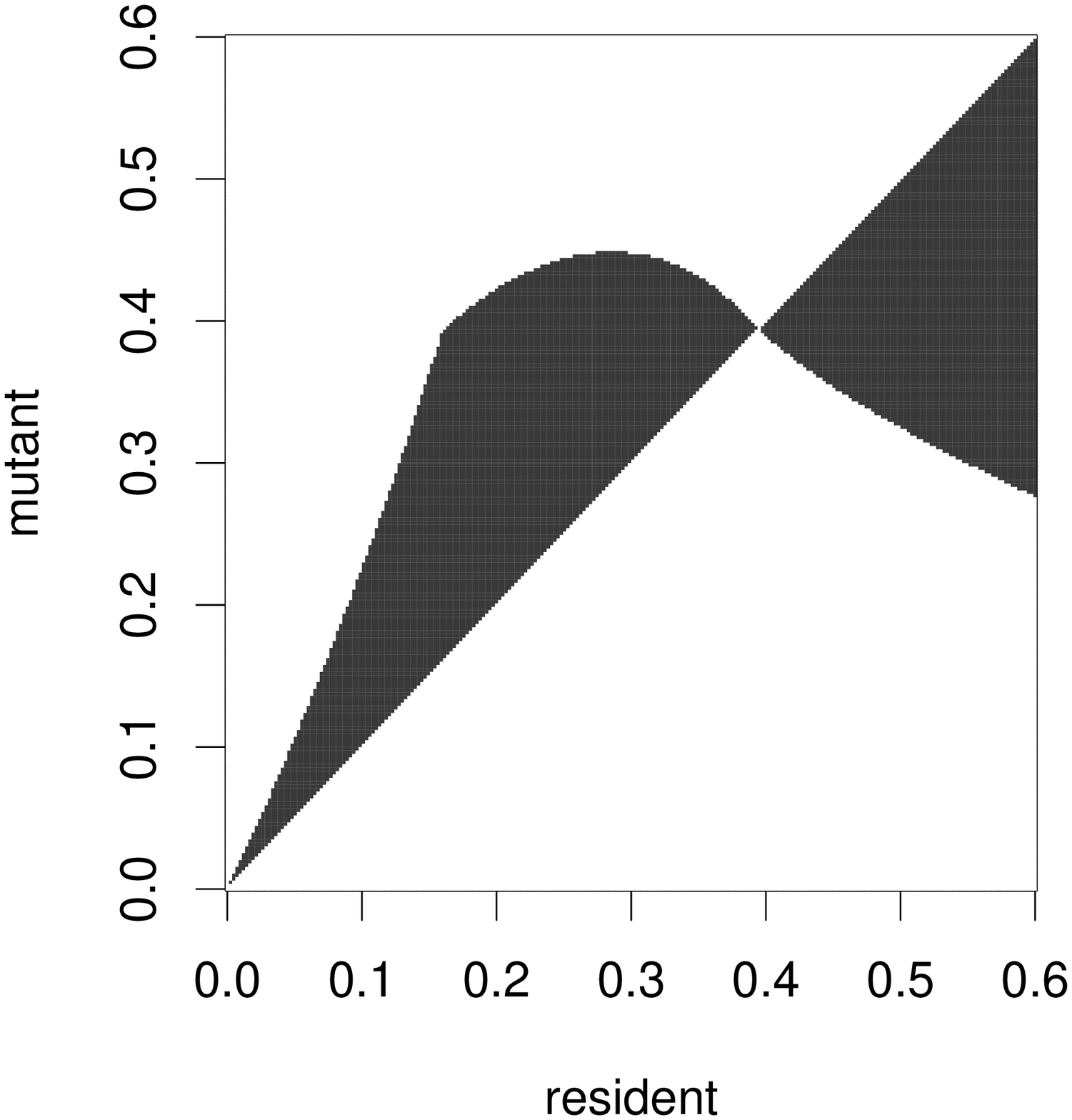}
	\end{center}
\caption{Pairwise Invasion Plot, PIP, with $b$ as trait. Dark region: Mutant can invade.  
	(a) $C(b)=\kappa_1\, b$ s.t. $C'(0)=\kappa_1>0$. 
	(b) $C(b)=\kappa_2\, b^2$, s.t.\ $C'(0)=0$. Model: Quiescence.
	Remaining parameters: 
	$\kappa_1=1/2$, $\kappa_2=1$, 
	$\beta      = 1$,
	$\mu_x        = 0.1$,
	$\mu_y        = 0.1$
}\label{PIPfig}
\end{figure}

We exemplify the last theorem. We choose 
the function $\A(b)$, paralleling inequality~(\ref{adapDynInequal}),  
$$\A'(b) = \zeta\,\frac{C'(b)}{b-C(b)}\, (1+\A(b)),\qquad \A(b_0)=A_0.$$
Here, we choose $\zeta>1$, such that $\A(b)$ satisfies the strict inequality (\ref{adapDynInequal}) in case that $b-C(b)>0$. We still have the freedom to 
select $C(b)$; we use two different functions, to exemplify the situations 
$C'(0)>0$ resp.\ $C'(0)=0$. 
\par\medskip 
 
\underline{Case 1:} $C(b)=\kappa b$, where $\kappa\in(0,1)$ (most natural case).\\
The solution of the ODE for $\A(b)$ reads
$$ \A(b) = (1+A_0)\,\,\left(\frac b {b_0}\right)^{\zeta\,\frac{\kappa}{1-\kappa}}-1$$
That is, we always have $\A(0)=-1$. Recall that our quiescent/seedbank model do not produce any function $\A(b)$, but there are restrictions. Let us take the quiescence model as a reference. In order to translate the 
$b$-dependent function $\A(b)$ into the original quiescence model with parameter function $A(\gamma)$, we need to map $b$ to $\gamma$ by the condition
$$ A(\gamma)=A(\gamma(b))=\A(b).$$
We know that $A(\gamma)$ has only a bounded range (appendix~\ref{quiescMonoRange}). 
We simply cut $\gamma(b)$ is $\A(b)$ leaves the range of 
$A(b)$: if $A(b)<A(0)$, we take $\gamma(b)=0$, and if $\A(b)>A(\infty)$, 
 then $\gamma=\infty$ (does not happen often). For the existence of an ESS with $b>0$ it is important that there is an interval for $b$, where we can satisfy $\A(b)=A(\gamma(b))$. \par\medskip 
 
\underline{Case 2:} $C(b)= \kappa\, b^2$.\\
In this case we obtain
$\A(b) = (1+A_0) \left(\frac {1-b_0} {1-b}\right)^{2\,\zeta\,\kappa}-1$. 
We can use $b_0=0$ as reference, and have 
 $$\A(b) = 
(1+A_0) \left(\frac {1} {1-b}\right)^{2\,\zeta\,\kappa}-1$$
such that $\A(0)=A_0$. Again, we use the quiescence model as a reference. Since $A(\gamma)$ is between $0$ and $A(\infty)\in(0,-1)$, $A_0$ is negative, but larger than $-1$. Hence, $\A(b)$ is increasing. If $\A$ exceeds $0$, 
there is no solution of $\A(b)=A(\gamma)$ any more.\par\medskip 

The Pairwise Invasion Plots (PIP) for these two cases in the quiescent model are presented in figure~\ref{PIPfig}. In both cases, these is a convergence stable ESS with cooperation ($b>0$).In line with the theorem above, trait $b=0$ is a convergent stable ESS in case~1 ($C'(0)>0$), 
while it is convergent unstable in case~2 ($C'(0)=0$). In the latter case, a small initial perturbation leads to an increasing amount of cooperation. While the cooperation is increasing, $\A(b)$ is increasing, the time scale of the resting state given by $\gamma$ decreases. Therefore, the higher the degree of cooperation, the longer becomes the quiescent state and conversely the shorter is the seedbank.
\par\medskip

\section{Discussion}
We investigate a model for quiescence or seedbanks, augmented by cooperation. We find that the system with weak frequency-dependent selection can be reduced to a generalized replicator equation. Comparing this equation with the classical replicator one, additional constants that express the time scale of the resting states of the respective types are found. Our main results state that in contrast to the classical results, for which cooperation is never (locally asymptotically) stable, the resting stages may lead to a stabilization and maintenance of cooperation in our models. This result is in line with simulation studies indicating that the co-evolution of age and time scales may foster cooperation~\citep{Wu2009,Rong2010,Liu2012,Rong2013}. Ultimately, the cooperating type needs to be slower than the defector in the quiescent model and faster in the seed bank model for this stabilizing effect to occur. Note that this observation is counter intuitive for quiescence, as the red queen assumption~\citep{Hauert2002,decaestecker2007} indicates that in general the species with the smallest generation time has an advantage and can out-compete the slower species (as we find in the seed bank model). However, in some cases of host-parasite coevolution, the host may outcompete the parasite despite having longer generation time, an effect termed as the red-king hypothesis~\citep{Bergstrom2003,Damore2011}. In this scenario, the hosts do set the stage and the possible outcomes and as such can be seen as potential defectors. The parasite, which reproduces several times within a host, is therefore forced to cooperate. Note that in the Red King theory, the parasites live within the hosts and thus hosts' and parasites' roles are not exchangeable. In the present study, the mechanisms that stabilize cooperation seem to be slightly different because cooperators and defectors are symmetric in most of their life-history traits aspects. \\
It is particularly interesting that the time scales of quiescence and seedbank should vary in opposite directions to promote and maintain cooperation. We explain this result by the difference timing in offspring production and when these offspring are present in the active population and thus under selection. In the quiescent model, all offspring are produced and enter the active population after the parent individual exists the quiescent state. In the seedbank model, offspring are produced when the parent is in the active population, but they become dormant and the offspring will germinate and enter the active population at different future time points. In other words, the competition between offspring of a given parent is immediate in the quiescent model while it is delayed in the seedbank. As cooperation is favored when its frequency is high enough so that cooperators do out-compete defectors, dormancy should be short for enough cooperators to be present at the same time, while this condition is fulfilled in the quiescence as several offspring directly enter the active population.  In order to be maintained, cooperators should evolve compared to defectors 1) longer quiescent times, or 2) shorter dormancy times.
In both models the evolution of quiescence or dormancy results in a higher fraction of the population that is resting, amplifying the beneficial effects of weak selection. If the defectors have a smaller resting fraction, their amplification factor is smaller than that of the cooperators, and cooperation can prevail. \par\medskip  
Using adaptive dynamics, we also show that in our populations with resting stages, 1) cooperation can appear, and 2) evolution leads to different optimal quiescent or dormancy strategies for cooperators and defectors. It is of interest to discuss the plausibility of these two events to happen and to be observed in natural systems. \par\medskip
We show that cooperation appears spontaneously if the resident trait without any cooperation ($b=0$) is, due \textit{e.g.} to a small perturbation, replaced by a trait with a little cooperation ($b>0$ but small) assuming the costs for cooperation being very small ($C'(0)=0$). The latter prerequisite could seem at first hard to satisfy in any real world system. In fact, some systems would meet these requirements. If bacteria use metabolic prudence~\citep{Schuster2013,Boyle2015}, they produce the public good at time periods at which other resources are the limiting factor (for growth) than the resources needed for producing the public good. In this case, the public good production does not superimpose a growth rate reduction, and thus the costs for the public good production vanish (or at least are very small). We speculate that this scenario at the initial state would allow the increase of public good production. The degree of cooperation is then expected to grow until a convergent stable ESS is reached. At the ESS, a certain public good production and a certain time scale for the resting stages are observed, and this combination would allow cooperation not only to develop but also to persist. Though, eventually the costs would increase later on and further evolution of the resting stages would occur. \par\medskip
Once cooperation can evolve and be maintained, defectors and cooperators do evolve different optimal quiescent or dormancy strategies. We are not aware of empirical observations or experimental evolution studies supporting our hypothesis. Future studies on bacteria experimental evolution could manipulate the rates of dormancy ~\citep{Beaumont2009,lennon2011} or quiescence by \textit{e.g.} manipulating the environmental variable such as resources over time or generating artificially seedbanking by freezing samples at regular time intervals. Such approach could be used in systems where bacteria are already used to study the evolution of cooperation~\citep{Schuster2013,Boyle2015}. In natural systems, we conjecture that the partners of symbiotic systems such as Angiosperms and Gymnosperms and their mycorhizal fungi~\citep{Strullu‐Derrien2018} may be good models to assess the existence of seedbanks and differential dormancy in cooperators or defectors. One would expect, for example, that the more cooperating strains of mycorrhizal fungi would show shorter dormancy times than the more defective strains. \par\medskip
Finally, we highlight that the present approach is based on the assumption that public good production and time scales are strongly intertwined, which may not always be true in all biological systems. Nevertheless, we propose a novel hypothesis for the evolution and maintenance of cooperation in an homogeneous system by the evolution of resting stages (quiescence or seedbanks).


\section*{Acknowledgements}
{\it This research is supported in part by Deutsche Forschungsgemeinschaft grants TE 809/1 (AT), and MU2339/2-2 from the Priority Program 1617 (JM).

\bibliographystyle{elsarticle-harv}
\bibliography{popGen,sleepyLit}

\newpage

\begin{appendix}

\section{System for singular perturbation theory -- Quiescence}

\subsection{Transformation of the system}
\label{QtransForm}
$\frac d {dt}\ln(N)$:
\begin{eqnarray*}
	\frac d {dt}\ln(N) 
	& = &
	\sum_{i=1}^2\bigg( (\beta_i + \omega \, g_i(\hx_1,\hx_2)- \alpha_i -\mu_{x,i} )\hx_i+\gamma_i \hy_i\bigg)\\
	& = &  \lambda^{(q)} + \omega \sum_{j=1}^2\bigg(   g_j(\hx_1,\hx_2)\tx_j +\gamma_j\ty_j   \bigg)
\end{eqnarray*}
$\dot \tx_1$:
\begin{eqnarray*}
\frac d {dt} \tx_1  &=& 
(\beta_1 + \omega g_1(\hx_1,\hx_2))\, \hx_1 - \alpha_1 \hx_1-\mu_{x,1} \hx_1+\gamma_1 \hy_1 -\hx_1\,\frac d {dt}\ln(N)\\
&=&(\beta_1 + \omega g_1(x_1,x_2))\, \tx_1 - \alpha_1 \tx_1
-\mu_{x,1} \tx_1+\gamma_1 \left(
\frac{-\beta_1  + \alpha_1 
	+\mu_{x,1} +\lambda^{(q)} }{\gamma_1}\,\,\tx_1
+\omega \ty_1
\right)
-\tx_1\,\frac d {dt}\ln(N)\\
&=&  \omega \, \tx_1\, 
\bigg\{ 
 g_1(\tx_1,\tx_2)\,  -
\sum_{i=1}^2\bigg(   g_i(\tx_1,\tx_2)\tx_i +\gamma_i\ty_i   \bigg)
\bigg\}+\omega\gamma_1\ty_1\\
&=&  \omega \, \tx_1\, 
\bigg\{ 
g_1(\tx_1,\tx_2)\,  -
g_1(\tx_1,\tx_2)\,\tx_1 - g_2(\tx_1,\tx_2)(1-\tx_1)\bigg\}
+\omega\bigg\{\gamma_1\,(1-\tx_1)\,\ty_1-\gamma_2\,\tx_1\,\ty_2\bigg\}
\\
&=&  \omega \,\left[ \tx_1\,(1-\tx_1)\, 
\bigg\{ 
g_1(\tx_1,\tx_2)- g_2(\tx_1,\tx_2)\bigg\}
+ \, \bigg\{
\gamma_1\,(1-\tx_1)\ty_1-\gamma_2\,\tx_1\,\ty_2\bigg\}\right].
\end{eqnarray*}
Let $\ol i=2$ if $i=1$ and {\it vice versa}. 
For symmetry reasons, 
$$
\frac d {dt} \tx_i 
=
\omega \,\left[ \tx_1\,\tx_2\, 
\bigg\{ 
g_i(\tx_1,\tx_2)- g_{\ol i}(\tx_1,\tx_2)\bigg\}
+ \, 
\bigg\{
\gamma_i\,\tx_{\ol i}\,\ty_i-\gamma_{\ol i}\, \tx_i \,\ty_{\ol i}\bigg\}\right].
$$
$\dot\ty_i$:
\begin{eqnarray*}
\frac d {dt} \hy_i  
&=&  \alpha_i \hx_i-\gamma_i \hy_i -\mu_{y,i} \hy_i 
-\hy_i\,\frac d {dt}\ln(N)
= 
 \alpha_i \tx_i
 -\bigg(\gamma_i+\mu_{y,i}+\frac d {dt}\ln(N)\bigg) \hy_i\\
 \Rightarrow\quad 
 \omega\frac{d}{dt} \ty_i &=& - 
  \frac{-\beta_i  + \alpha_i 
 	+\mu_{x,i} +\lambda^{(q)} }{\gamma_i}\,\,\frac d {dt}\,\hat x_i
 + \alpha_i \tx_i
 -\bigg(\gamma_i+\mu_{y,i}+\frac d {dt}\ln(N)\bigg) \hy_i.
\end{eqnarray*}
 Then, 
\begin{eqnarray*}
	\omega \frac d {dt}\ty_i 
	&=& -
	\frac{-\beta_i  + \alpha_i 
		+\mu_{x,i} +\lambda^{(q)} }{\gamma_i}\,\,
	\left( 
	 \omega \,\left[ \tx_1\,\tx_2\, 
	\bigg\{ 
	g_i(\tx_1,\tx_2)- g_{\ol i}(\tx_1,\tx_2)\bigg\}
	+ \, 
	\bigg\{
	\gamma_i\,\tx_{\ol i}\,\ty_i-\gamma_{\ol i}\, \tx_i \,\ty_{\ol i}\bigg\}\right]
	   \right)\\
&& +\alpha_i \tx_i
-\bigg(\gamma_i+\mu_{y,i}+\frac d {dt}\ln(N)\bigg) 
\left( 
\frac{-\beta_i  + \alpha_i 
	+\mu_{x,i} +\lambda^{(q)} }{\gamma_i}\,\,\hat x_i
+\omega  \ty_i
\right)\\
	&=& -
\frac{-\beta_i  + \alpha_i 
	+\mu_{x,i} +\lambda^{(q)} }{\gamma_i}\,\,
\left( 
\omega \,\left[ \tx_1\,\tx_2\, 
\bigg\{ 
g_i(\tx_1,\tx_2)- g_{\ol i}(\tx_1,\tx_2)\bigg\}
+ \, 
\bigg\{
\gamma_i\,\tx_{\ol i}\,\ty_i-\gamma_{\ol i}\, \tx_i \,\ty_{\ol i}\bigg\}\right]
\right)\\
&& -\frac{\tx_i}{\gamma_i}\,\,
\left(
-\alpha_i \gamma_i
+(\gamma_i+\mu_{y,i}+\lambda^{(q)})
(-\beta_i  + \alpha_i 
+\mu_{x,i} +\lambda^{(q)})
\right)\\
&&-  \omega \,
\left(
\frac{-\beta_i  + \alpha_i 
	+\mu_{x,i} +\lambda^{(q)} }{\gamma_i}\,\,\right)
\hx_i\,\sum_{j=1}^2\bigg(   g_j(\hx_1,\hx_2)\tx_j +\gamma_j\ty_j   \bigg)  
-\omega \bigg(\gamma_i+\mu_{y,i}+\frac d {dt}\ln(N)\bigg) 
 \ty_i
\end{eqnarray*}
With the characteristic equation for $\lambda^{(q)}$
$$
(\lambda_i^{(q)})^2 - (\beta_i-\alpha_1-\mu_{x,i}-\gamma_i-\mu_{y,i}) 
(\lambda_i^{(q)}) - 
(\beta_i-\alpha_1-\mu_{x,i})(\gamma_i+\mu_{y,i})
-\alpha_i\gamma_i
=0
$$
the ${\cal O}(\omega^0)$-term in this equation becomes zero. 
Dividing by $\omega$ yields 
\begin{eqnarray*}
	 \frac d {dt}\ty_i 
	&=& 
-\,\,\frac{-\beta_i  + \alpha_i 
	+\mu_{x,i} +\lambda^{(q)} }{\gamma_i}\,\,
\bigg( 
 \, \tx_1\,\tx_2\, 
\bigg\{ 
g_i(\tx_1,\tx_2)- g_{\ol i}(\tx_1,\tx_2)\bigg\}
+\tx_i\,\sum_{j=1}^2\bigg(   g_j(\tx_1,\tx_2)\tx_j +\gamma_j\ty_j \bigg)\\
&&
+ \, 
\bigg\{
\gamma_i\,\tx_{\ol i}\,\ty_i-\gamma_{\ol i}\, \tx_i \,\ty_{\ol i}\bigg\}
\bigg) 
- \bigg(\gamma_i+\mu_{y,i}+\frac d {dt}\ln(N)\bigg) 
\ty_i\\
	&=& 
\frac{-\beta_i  + \alpha_i 
	+\mu_{x,i} +\lambda^{(q)} }{\gamma_i}\,\,
\bigg( 
\,\bigg[ \tx_i\,
g_i(\tx_1,\tx_2)
+\tx_i\,(\gamma_1\ty_1+\gamma_2\ty_2)
+ \, 
\bigg\{
\gamma_i\,\tx_{\ol i}\,\ty_i-\gamma_{\ol i}\, \tx_i \,\ty_{\ol i}\bigg\}\bigg]
\bigg) \\
&&- \bigg(\gamma_i+\mu_{y,i}+\frac d {dt}\ln(N)\bigg) 
\ty_i\\
	&=& -
	\,\frac{-\beta_i  + \alpha_i 
	+\mu_{x,i} +\lambda^{(q)} }{\gamma_i}\,\,
\,\bigg( \tx_i\,g_i(\tx_1,\tx_2)+\,\gamma_i\,\ty_i\bigg) 
- \bigg(\gamma_i+\mu_{y,i}+\frac d {dt}\ln(N)\bigg) 
\ty_i\\
	&=& 
	-\frac{-\beta_i  + \alpha_i 
	+\mu_{x,i} +\lambda^{(q)} }{\gamma_i}\,\,
\,\,\, \tx_i\,g_i(\tx_1,\tx_2)
+  
 \bigg(\beta_i  - \alpha_i 
-\mu_{x,i} -\lambda^{(q)}-\gamma_i-\mu_{y,i}-\frac d {dt}\ln(N)\bigg) 
\ty_i
\end{eqnarray*}

\subsection{Quiescence, fast system}
\label{qInequlaities}

Take $\omega$ to zero. $\tx_i$ will not change any more, and 
$\frac {d}{dt}\ln(N)=\lambda^{(q)}$. Then, $\ty_i$ satisfy linear 
equations. The factor in front of $\ty_i$ is negative:
Let 
$$
A = \beta_i  - \alpha_i 
-\mu_{x,i} -\gamma_i-\mu_{y,i}.
$$
We show that $A<2\lambda^{(q)}$, that is, $A/2<\lambda^{(q)}$. 
\begin{prop} We have the inequalities
\begin{eqnarray}
2\, \lambda^{(q)} &>&  \beta_i  - \alpha_i -\mu_{x,i} -\gamma_i-\mu_{y,i} 
\label{inequQuiesc1}\\
 \lambda^{(q)} &>& \beta_i  - \alpha_i -\mu_{x,i}.\label{inequQuiesc2}
\end{eqnarray}
If $\lambda^{(q)}>0$, we have additionally
\begin{eqnarray}
\lambda^{(q)} &<&  \beta_i  - \mu_{x,i}.\label{inequQuiesc3}
\end{eqnarray}
\end{prop}
{\bf Proof: } 
Recall that 
\begin{eqnarray*}
p_q(\lambda) &=& 	\lambda^2 - (\beta_i-\alpha_i-\mu_{x,i}-\gamma_i-\mu_{y,i})\lambda - 
	(\beta_i-\alpha_i-\mu_{x,i})(\gamma_i+\mu_{y,i})
	-\alpha_i\gamma_i\\
	&=& 
-(\beta_i-\alpha_i-\mu_{x,i}-\lambda)
(\lambda+\gamma_i+\mu_{y,i})
	-\alpha_i\gamma_i.
\end{eqnarray*}
If we plug in $A/2$ into the 
characteristic equation for the quiescence model, we find 
\begin{eqnarray*}
p_q(A/2) &=& A^2/4 - A^2/2 - 
(\beta_i-\alpha_1-\mu_{x,i})(\gamma_i+\mu_{y,i})-\alpha_i\gamma_i \\
&=& \frac {-1} 4\,\,( (\beta_i  - \alpha_i 
-\mu_{x,i} -(\gamma_i+\mu_{y,i}))^2+4 (\beta_i-\alpha_1-\mu_{x,i})(\gamma_i+\mu_{y,i})) -\alpha_i\gamma_i \\
&=& \frac {-1} 4\,\,( \beta_i  - \alpha_i 
-\mu_{x,i} +\gamma_i+\mu_{y,i})^2 -\alpha_i\gamma_i
< 0.
\end{eqnarray*}	
As $\lambda^{(q)}$ is the larger root (which we know to be real) 
of the polynomial, and the squared term has a positive sign, 
$A/2<\lambda^{(q)}$. Hence $A-2\lambda^{(a)}<0$.\par\medskip 

Second, we show the inequality
$$\beta_i  - \alpha_i -\mu_{x,i} -\lambda^{(q)}<0 $$ 
 We use the same approach as above, 
and inspect $p_q(.)$ at $\beta_i  - \alpha_i -\mu_{x,i}$.
\begin{eqnarray*}
	p_q(\beta_i  - \alpha_i -\mu_{x,i}) &=& 
	(\beta_i  - \alpha_i -\mu_{x,i})^2 
	-(\beta_i  - \alpha_i -\mu_{x,i})\,(\beta_i  - \alpha_i 
	-\mu_{x,i} -\gamma_i-\mu_{y,i})\\
&&	- 
	(\beta_i-\alpha_1-\mu_{x,i})(\gamma_i+\mu_{y,i})
	-\alpha_i\gamma_i\\
	&=&-\alpha_i\gamma_i<0.
\end{eqnarray*}
As above, the inequality is an immediate consequence of 
$p_q(\beta_i  - \alpha_i -\mu_{x,i})<0$.\\
In order to show the third inequality, we first note 
that $\lambda^{(q)}>0$ implies $\beta_i-\mu_{x,i}>0$: Therefore, we start with 
the observation that the roots of $p_q(\lambda)$ read
$$\lambda_\pm = 
\frac 1 2 \bigg[
\beta_i-\alpha_1-\mu_{x,i}-\gamma_i-\mu_{y,i}
\pm\sqrt{
	(\beta_i-\alpha_1-\mu_{x,i}+\gamma_i+\mu_{y,i})^2
	+4\alpha_i\gamma_i
}
\bigg].
$$
If $\beta_i-\alpha_1-\mu_{x,i}+\gamma_i+\mu_{y,i}>0$, 
$$\lambda_-<
\frac 1 2 \bigg[
\beta_i-\alpha_1-\mu_{x,i}-\gamma_i-\mu_{y,i}
-
	(\beta_i-\alpha_1-\mu_{x,i}+\gamma_i+\mu_{y,i})
\bigg]<-\gamma_i-\mu_{y,i}
<0$$
 and if 
 $\beta_i-\alpha_1-\mu_{x,i}+\gamma_i+\mu_{y,i}<0$ (and hence 
 $\beta_i-\alpha_1-\mu_{x,i}<-\gamma_i-\mu_{y,i}$
 ), 
 $$\lambda_-<
 \frac 1 2 \bigg[
 \beta_i-\alpha_1-\mu_{x,i}-\gamma_i-\mu_{y,i}
 +
 (\beta_i-\alpha_1-\mu_{x,i}+\gamma_i+\mu_{y,i})
 \bigg] < \beta_i-\alpha_1-\mu_{x,i}
 <0.$$
 In any case, $\lambda_-<0$. Hence $\lambda_i^{(q)}>0$ implies $p_q(0)<0$, and 
 \begin{eqnarray*}
 0&>&
 -(\beta_i-\alpha_1-\mu_{x,i})
 (\gamma_i+\mu_{y,i})
 -\alpha_i\gamma_i 
 = -(\beta_i-\mu_{x,i})(\gamma_i+\mu_{y,i})+\alpha\mu_{y,i}
 \end{eqnarray*}
which in turn forces $\beta_i-\mu_{x,i}>0$.\\
To show $\lambda^{(q)}<\beta_i-\mu_{x,i}$, we 
inspect $p_q(\lambda)$ at $\lambda=\beta_i-\mu_{x,i}$, and find 
\begin{eqnarray*}
p_q(\beta_i-\mu_{x,i}) &=& 
	\alpha_i(\beta_i-\mu_{x,i}+\mu_{y,i})>0
\end{eqnarray*}
which implies the desired inequality.
\par\qed\par\medskip

\subsection{Quiescence: dependence of $A_i^{(q)}$ on time scale}
\label{quiescMonoRange}
\subsubsection{$A$'s monotonicity}
The time scale of the quiescent state is given by $\gamma$ and $\alpha$. 
In the following, we assume that the exponential growth rate 
$\lambda^{(q)}$ is determined for a certain parameter set; 
we then vary the time scale of 
the quiescent state in varying $\gamma$, and adapt $\alpha$ such that the exponential growth does not change. In this way, we define $\alpha=\alpha(\gamma)$, and discuss the dependency of 
\begin{eqnarray*}
	A^{(q)}(\gamma) =-\,\, \frac{\beta-\alpha(\gamma)-\mu_{x}-\lambda^{(q)}}
	{\beta-\alpha(\gamma)-\mu_{x}-\gamma-\mu_{y}-2\lambda^{(q)}}
\end{eqnarray*}

\begin{prop}\label{monotoneQuiesc}
	Assume that there are $\alpha,\gamma>0$, such that 
$\lambda^{(q)}$ is the positive root of $p_q(\lambda)$ (the growth rate). Assume 
furthermore that $\lambda^{(q)}>0$. Then, for all $\gamma>0$, $\alpha(\gamma)>0$. Furthermore, $	A^{(q)}(\gamma)$ is monotonously decreasing.
\end{prop}
{\bf Proof: } As the (larger) root of $p_q(\lambda)$ does not change, we can solve  $p_q(\lambda^{(q)})=0$ for $\alpha$ and find 
\begin{eqnarray*}
\alpha &=& \alpha(\gamma) = -\,\,\frac
{(\lambda^{(q)}-(\beta-\mu_{x}))\,(\lambda^{(q)}+(\gamma+\mu_{y}))}
{\lambda^{(q)}+\mu_{y}}
\end{eqnarray*}
$\alpha(\gamma)$ is linear in $\gamma$, 
and cannot change sign for $\gamma>0$. 
As we assume that there is $(\gamma,\alpha)\in\R_+^2$ that yield the eigenvalue $\lambda^{(q)}$, $\alpha(\gamma)$ is positive for all $\gamma>0$, and increasing. In particular, we have for the derivative of $\alpha$ w.r.t.~$\gamma$ 
$$ 
\alpha'(\gamma) = -\,\,\frac
{(\lambda^{(q)}-(\beta-\mu_{x}))}
{\lambda^{(q)}+\mu_{y}}\qquad\Rightarrow\qquad
\alpha(\gamma) = \alpha'(\gamma)\, (\lambda^{(q)}+\gamma+\mu_{y}). $$
We rewrite $A^{(q)}$ as
\begin{eqnarray*}
&&	A^{(q)} 
= -\,\,\frac{\beta-\alpha-\mu_{x}-\lambda^{(q)}}
{ (\beta-\alpha-\mu_{x}-\lambda^{(q)})\,-(\gamma+\mu_{y}+\lambda^{(q)})}
= \frac{\gamma+\mu_{y}+\lambda^{(q)}}
{ (2\,\lambda^{(q)}-\beta+\mu_x+\mu_y) +\alpha+\gamma} - 1.
\end{eqnarray*}
Inequality (\ref{inequQuiesc1}) implies that the denominator 
is always positive, such that $A^{(q)}$ is for all $\gamma>0$ well defined. 
In order to discuss the dependence of $A^{(q)}$ on $\gamma$ (note that $\alpha=\alpha(\gamma)$ 
and $\lambda^{(q)}-\beta+\mu_x<0$ due to inequality~(\ref{inequQuiesc3})) 
we take the derivative,
\begin{eqnarray*}
\frac d {d\gamma}A^{(q)}(\gamma) &=& 
\frac{\lambda^{(q)}-\beta+\mu_x+\alpha(\gamma)-\alpha'(\gamma)(\gamma+\mu_{y}+\lambda^{(q)})}{((2\,\lambda^{(q)}-\beta+\mu_x+\mu_y) +\alpha+\gamma)^2}
= 
\frac{\lambda^{(q)}-\beta+\mu_x}{((2\,\lambda^{(q)}-\beta+\mu_x+\mu_y) +\alpha+\gamma)^2}<0.
\end{eqnarray*}
\par\qed

\subsubsection{Values for $A$}
Which values does $A$ assume? First of all, 
$$\alpha(0) 
= 
-\,\,\frac
{(\lambda^{(q)}-(\beta-\mu_{x}))\,(\lambda^{(q)}+\gamma+\mu_{y})}
{\lambda^{(q)}+\mu_{y}}\bigg|_{\gamma=0} 
= -(\lambda^{(q)}-(\beta-\mu_{x}))
$$
and hence 
\begin{eqnarray*}
A^{(q)}(0) 
&=& 
\frac{\gamma+\mu_{y}+\lambda^{(q)}}
{ (2\,\lambda^{(q)}-\beta+\mu_x+\mu_y) +\alpha+\gamma}\bigg|_{\gamma=0} - 1 
= 
\frac{\mu_{y}+\lambda^{(q)}}
{ (2\,\lambda^{(q)}-\beta+\mu_x+\mu_y)-(\lambda^{(q)}-\beta+\mu_{x})} - 1 = 0.
\end{eqnarray*}
Next we note that 
$$\lim_{\gamma\rightarrow\infty} \frac{\alpha(\gamma)}{\gamma} 
= -\frac{(\lambda^{(q)}-\beta+\mu_x)}{\lambda^{(q)}+\mu_y}
$$
and hence 
$$ 
\lim_{\gamma\rightarrow\infty} A(\gamma) 
= 
\frac{1}{1-\frac{(\lambda^{(q)}-\beta+\mu_x)}{\lambda^{(q)}+\mu_y}}-1
= 
\frac{\lambda^{(q)}+\mu_y}{\beta-\mu_x+\mu_y} - 1 
= 
-\frac{\beta-\mu_x-\lambda^{(q)}}{\beta-\mu_x+\mu_y}.
$$
Note that $\lambda^{(q)}>0$ implies that $\beta-\mu_x>\lambda^{(q)}>0$, such that this expression is well defined and negative for non-decreasing population size. 
Furthermore, for $\lambda^{(q)}>0$, we have 
$\frac{\lambda^{(q)}+\mu_y}{\beta-\mu_x+\mu_y}>0$, such that $A(\infty)>-1$. 
\\
Thus, $A(\gamma)$ is monotonously decreasing from $A(0)=0$ 
to $A(\infty)\in(0,-1)$.

\newpage

\section{System for singular perturbation theory -- seed banking}

\subsection{Transformation of the system}
\label{StransForm}
$\frac{d}{dt}ln(N)$:
\begin{eqnarray*}
\frac{d}{dt}ln(N)
&=&\sum_{i=1}^{2}(-\mu_{x,i}\,\tx_{i}+\gamma_{i}\hy_{i} )
=\sum_{i=1}^{2}-\bigg[\mu_{x,i}\,\tx_{i}+\gamma_{i}
\bigg(\frac{(\mu_{x,i}+\lambda^{(s)})\tx_{i}}{\gamma_{i}}+\omega\ty_{i}\bigg)\bigg]\\
&=&
\lambda^{(s)}+ \omega\,\sum_{i=1}^{2}\gamma_{i}\,\ty_{i}
\end{eqnarray*}

Let i in $\{ 1,2 \}$, for symmetry reasons we have : 

$\dot\tx_i$:
\begin{eqnarray*}
\frac{d}{dt}\tx_{i}&=&-\tx_{i}(\mu_{x,i}+\frac{d}{dt}ln(N))+\gamma_{i}\hy_{i}
= -\tx_{i}\left(\mu_{x,i}+\lambda^{(s)}+ \sum_{i=1}^{2}\omega\gamma_{i}\ty_{i}\right)+\gamma_{i}\left(\frac{(\mu_{x,i}+\lambda^{(s)})\tx_{i}}{\gamma_{i}}+\omega\ty_{i}\right)\\
&=&\omega\left(\gamma_{i}\ty_{i}-\tx_{i}\sum_{j=1}^{2}\gamma_{j}\ty_{j}\right)
\end{eqnarray*}

$\dot\ty_i$:
\begin{eqnarray*}
\omega\,\frac{d}{dt}\ty_{i}
&=&\frac{d}{dt}\hy_{i} -  \frac{\mu_{x,i}+\lambda^{(s)}}{\gamma_{i}} \frac{d}{dt}\tx_{i}\\
&=&
\bigg((\beta_i + \omega g_i(\tx_1,\tx_2))\, \tx_i - \mu_{y,i} \hy_i -\gamma_i \hy_i
-\hy_i\,\frac d {dt}\ln(N) \bigg)\\
&& - \omega\,\left( \frac{\mu_{x,i}+\lambda^{(s)}}{\gamma_{i}}\right)
 \left(\gamma_{i}\ty_{i}-\tx_{i}\sum_{j=1}^{2}\gamma_{j}\ty_{j}\right)\\
&=&
 (\beta_i + \omega g_i(\tx_1,\tx_2)\,)\, \tx_i 
 - \left(\mu_{y,i} +\gamma_i  + \frac d {dt}\ln(N)   \right)
 \left(  \frac{\mu_{x,i}+\lambda^{(s)} }{\gamma_i  }\,\, \tx_i +\omega\ty_i\right)\\
&& - \omega\,\left( \frac{\mu_{x,i}+\lambda^{(s)}}{\gamma_{i}}\right)
 \left(\gamma_{i}\ty_{i}-\tx_{i}\sum_{j=1}^{2}\gamma_{j}\ty_{j}\right)\\
 &=&
\left[\beta_i\, \tx_i 
 - \left(\mu_{y,i} +\gamma_i  +\lambda^{(s)}+ \omega\,\sum_{j=1}^{2}\gamma_{j}\,\ty_{j}   \right)
 \left(  \frac{\mu_{x,i}+\lambda^{(s)} }{\gamma_i  }\,\right)\, \tx_i\right]\\
&& +\omega\left[
g_i(\tx_1,\tx_2)\, \tx_i
- \left(\mu_{y,i} +\gamma_i  + \frac d {dt}\ln(N)   \right)\,\ty_i
-\left( \frac{\mu_{x,i}+\lambda^{(s)}}{\gamma_{i}}\right)
\left(\gamma_{i}\ty_{i}-\tx_{i}\sum_{j=1}^{2}\gamma_{j}\ty_{j}\right)
\right]\\
 &=&
 \frac 1 {\gamma_i}\,\,
\left[\gamma_i\,\beta_i\, 
- \left(\mu_{y,i} +\gamma_i  +\lambda^{(s)}   \right)
\left(\mu_{x,i}+\lambda^{(s)} \,\right)\, \right]\,\, \tx_i\\
&& +\omega\left[
g_i(\tx_1,\tx_2)\, \tx_i
- \left(\mu_{y,i} +\gamma_i  + \frac d {dt}\ln(N)   \right)\,\ty_i
-\,(\mu_{x,i}+\lambda^{(s)})\,\ty_{i}
\right]
\end{eqnarray*}
The zero order term in $\omega$ becomes zero, as the term in the bracket equals the characteristic polynomial 
for the seedbank model. If we divide by $\omega$ we obtain
\begin{eqnarray*}
\frac{d}{dt}\ty_{i}
	&=&
g_i(\tx_1,\tx_2)\, \tx_i
- \left(\mu_{y,i} +\gamma_i +\mu_{x,i}+\lambda^{(s)} + \frac d {dt}\ln(N)   \right)\,\ty_i.
\end{eqnarray*}

\subsection{Seedbank, fast system}

\begin{prop}\label{qInequlaitiesB}
	 If we are in the fast time scale, taking $\omega\rightarrow 0$. Then, $\dot\tx_i=0$, 
$\frac d{dt}ln(N)=\lambda^{(s)}$, and the equation for 
$\ty_i$ becomes linear and  the constants of this linear equation are negative.
\end{prop}

{\bf Proof: }
We find 
\begin{eqnarray*}
\lim_{\omega\rightarrow0} \frac{d}{dt}\tx_{i}&=&\lim_{\omega\rightarrow0} \omega\left(\gamma_{i}\ty_{i}-\tx_{i}\sum_{j=1}^{2}\gamma_{j}\ty_{j}\right)=0,\qquad 
%
\lim_{\omega\rightarrow0} \frac{d}{dt}ln(N)=\lim_{\omega\rightarrow0} \lambda^{(s)}+ \omega\sum_{i=1}^{2}\gamma_{i}\,\ty_{i}=\lambda^{(s)}.
\end{eqnarray*}
Take $\omega$ to zero. $\tx_i$ will not change anymore, and 
$\frac {d}{dt}\ln(N)=\lambda^{(s)}$. Therefore, $\ty_i$ satisfies a linear equation. Since $\lambda^{(s)}$ is a positive root and all rates are positive we have :
\begin{eqnarray*}
-\left(\mu_{y,i} +\gamma_i +\mu_{x,i}+\lambda^{(s)} + \frac d {dt}\ln(N)   \right)\, < 0
\end{eqnarray*}

\par\qed

\subsection{Seedbank: dependency of $A_i^{(s)}$ on time scale}
\label{seedbankMonoRange}
\subsubsection{$A$'s monotonicity}

\label{monotoneSeed}

\begin{prop}
	Assume that there are $\mu_x,\gamma>0$, such that 
	$\lambda^{(s)}$ is the positive root of $p_s(\lambda)$ (the growth rate). Assume 
	furthermore that $\lambda^{(s)}>0$. Then, 
	 $A^{(s')}(\gamma)$ is monotonously increasing.
\end{prop}
{\bf Proof: }
Recall that $\mu_x$ is a function of $\gamma_i$ to ensure that 
the growth rate $\lambda$ is independent on the time scale of 
the system as defined by $\mu_{x,i}$ and $\gamma_i$. We use
$$ A^{(s')}(\gamma) = -\, \frac{\mu_x(\gamma)+\mu_y+2\lambda}
{\mu_x(\gamma)+\gamma+\mu_y+2\lambda} 
= 
-1 +  \frac{\gamma}
{\mu_x(\gamma)+\gamma+\mu_y+2\lambda} .
$$
The derivative of $\mu_x(\gamma)$ can be obtained from 
the equation for $\lambda$, 
\begin{eqnarray*}
\mu_x(\gamma) &=& 
\frac{\beta\gamma-\lambda^2-(\gamma+\mu_y)\lambda}
{\gamma+\mu_y+\lambda} 
= 
\frac{\beta\gamma}
{\gamma+\mu_y+\lambda}
-\lambda
\\
\mu_x'(\gamma) &=& 
\frac{\beta(\mu_y+\lambda)}
{(\gamma+\mu_y+\lambda)^2}
= \frac \beta {\gamma+\mu_y+\lambda}- \frac{\beta\gamma}{(\gamma+\mu_y+\lambda)^2} .
\end{eqnarray*}
Therefore, 
$\gamma\mu_x'(\gamma)-\mu_x(\gamma) 
= 
\lambda -\frac{\beta\gamma^2}{(\gamma+\mu_y+\lambda)^2}$, and
\begin{eqnarray*}
	\frac d{d\gamma} A^{(s')}(\gamma)
	&=&
	-\frac {-1}{\mu_x(\gamma)+\gamma+\mu_y+2\lambda}
	-\frac{\gamma\,(1+\mu_x'(\gamma))}{(\mu_x(\gamma)+\gamma+\mu_y+2\lambda)^2}\\
&=& 
	-\,\frac {\gamma\,\mu_x'(\gamma)-\mu_x(\gamma)-\mu_y-2\lambda}
	{(\mu_x(\gamma)+\gamma+\mu_y+2\lambda)^2}
=  \frac {\frac{\beta\gamma^2}{(\gamma+\mu_y+\lambda)^2}+\mu_x(\gamma)+\mu_y+\lambda}
{(\mu_x(\gamma)+\gamma+\mu_y+2\lambda)^2} >0 0.
\end{eqnarray*}

\subsubsection{Values for $A$}

We find directly that 
%
%
%
$$ A^{(s')}(0) = -\, \frac{\mu_x(0)+\mu_y+2\lambda}
{\mu_x(0)+0+\mu_y+2\lambda} =-1.
$$
Furthermore, from 
$\lim_{\gamma\rightarrow\infty} \mu_x(\gamma) = \beta - \lambda$
we conclude that 
\begin{eqnarray*}
\lim_{\gamma\rightarrow\infty} A^{(s')}(\gamma) &=& 0.
\end{eqnarray*}
Now, let us assume that there is $\gamma_0,\mu_{x,0}>$ such that the corresponding $\lambda^{s)}>0$. We fix this growth rate $\lambda^{s)}>0$, vary $\gamma$ and adapt 
$\mu_{x,0}$ accordingly to meet the condition that $\lambda^{(s)}$ is constant in $\gamma$. 
From $p_s(\lambda^{(s)})=0$ we conclude that 
$$\mu_x = 
\frac{\beta\gamma-(\lambda^{(s)})^2 -(\gamma+\mu_y)\lambda^{(s)}}{\lambda^{(s)}+\gamma+\mu_y}.
$$
The denominator always is positive. The numerator, however, 
changes sign if $\gamma$ drops below 
$$\widehat\gamma_0 = \frac{(\lambda^{(s)})^2 -(\gamma+\mu_y)\lambda^{(s)}}{\beta}.$$

\end{appendix}

\end{document}